# Routing in Wireless Adhoc Networks: A New Horizon

Mano Yadav, Vinay Rishiwal and K. V. Arya

**Abstract**— A lot of work has been done on routing protocols for mobile ad hoc networks, but still standardization of them requires some more issues less addressed by the existing routing protocols. In this paper a new paradigm of maintaining multiple connections in adhoc routing protocols has been highlighted which may be crucial for efficient routing in mobile ad hoc networks. The problem of multiple connections has been hardly worked on in adhoc networks. In this paper the solution of route maintenance if nodes are maintaining multiple connections has been proposed. This idea not only helps to solve the multiple connections problem, but also take care of proper bandwidth distribution to different connections as per different traffic types. Study has been incorporated on existing AODV with changes. Simulation studies have been performed over packet delivery ratio, throughput and message overheads. Results show that the proposed solution for multiple connections is efficient and worth implementing in existing as well as new protocols.

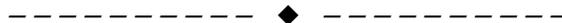

## 1 INTRODUCTION

An adhoc wireless network consists of a set of mobile nodes (hosts) that are connected by wireless links [1]. The network topology (physical connectivity of network) in such a network may keep changing randomly. Routing protocols that find a path to be followed by data packets from a source node to a destination node used in traditional wired networks can not be directly applied in adhoc wireless networks due to their highly dynamic topology, absence of established infrastructure for centralized administration, bandwidth constrained wireless links, and resource (energy) constrained nodes. A variety of routing protocols for adhoc wireless networks has been proposed in the recent past. Routing protocols in Mobile ad hoc networks can be classified in to two broad categories: proactive or table driven routing protocols, reactive or on demand routing protocols.

The proactive or table driven protocols attempt to find a route continuously and maintain routing information from each node to every other node within the network, so that whenever data is needed to be transmitted between two nodes, the route is already there. The nodes are required to maintain consistent up-to-date routing information in one or more tables. DSDV [2], WRP [3], CGSR [4] etc are the protocols of this category. In reactive approach routes are created only when desired by the source node. When a node requires a route to other node, it initiates a route discovery process with in the network. All permutations are examined and best possible route is established. The route is maintained by some route maintenance procedure until either the destination becomes inaccessible or until the route is no longer desired. AODV [5], DSR [6], LMR [7], TORA [8], and ABR [9] etc are reactive protocols.

This paper addresses the problem of routing in mobile ad hoc network with multiple connections. Since mobile nodes in mobile ad hoc network can move randomly the topology may change arbitrarily and frequently at unpredictable times. Transmission and reception parameters may also impact the topology. So it is very difficult to find and maintain an optimal route. The routing algorithm must react quickly to topologi-

- *Mano yadav is with the ITS Engineering College, Gretaer Noida, UP, India.*
- *Vinay Rishiwal is with the Department of CS & IT, IET, MJP Rohilkhand Ubniversity, Bareilly, UP, India.*
- *Prof. K.V Arya is with the ABV-Indian Institute of Information Technology, Gwalior, MP, India.*

cal changes. Nodes in Mobile ad hoc network communicate over wireless links. Therefore efficient use of bandwidth is a major issue in mobile ad hoc networks because wireless links are much lower in capacity. As the nodes are dynamic the number of nodes in certain bandwidth is always changing thus the capacity of links also keeps changing. The issue of maintaining multiple connections through wireless nodes is not being paid attention in the existing literature. This issue has been pointed out in this paper and a new routing scheme has been proposed for handling multiple connections in adhoc routing schemes.

Rest of the paper is organized as follows. The Section 2 deals with the concept of multiple connection issue. A new algorithm has been proposed for multiple connection maintenance in section 3. Section 4 provides simulation details and results has been discussed in section 5.

## 2 MULTIPLE CONNECTIONS IN MANETs

Most of the existing routing protocols have been developed considering single connection only. But a user may be involved in multiple connections for example downloading a file using one connection while talking to someone using another connection. This scenario is different from single connection routing as there may be several issues which do not arise during single connection routing. Let us consider figure 1, where every black dot represents a mobile node and its surrounding circle represents the coverage area of the mobile node. As shown in fig 1(a) node S is communicating with two parties D1 and D2 simultaneously. A route is maintained from S to D1 through neighbor N1 while route from S to D2 passes through neighbor N2. Now S moves and topology changes as shown in figure 1(b), and all the routes have to be re-established accordingly. Most of the protocols provide a solution to the problem of route re-establishment considering single connection only. Theoretically the same solution is applicable if a node is communicating with multiple parties simultaneously but there are several issues which may arise in this scenario as discussed below.

### 2.1 Parallel versus Serial Selection

As mobile node moves from current location to new one an important consideration is whether or not these routes can be re-established simultaneously. Simultaneous route re-establishment depends on the processing capabilities of the mobile nodes. If nodes are not capable to re-establish the routes simultaneously then routes have to be re-established serially. Now question arises which route should be re-established first? If the routes are re-established serially with out considering the traffic type, the real time traffic may suffer longer delays.

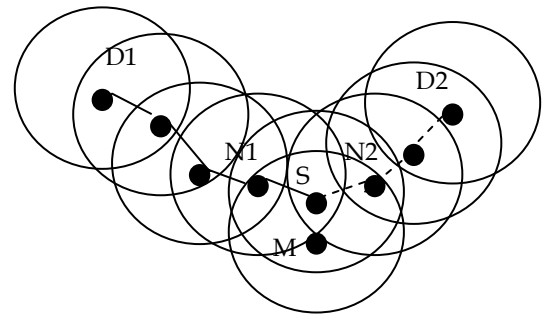

(a) Initial routes of two simultaneous connections of node S.

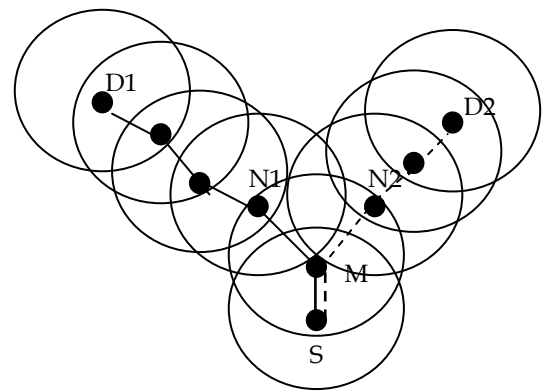

(b) Re-established routes when S moves to new location.

Fig 1: Multiple Connections of a Single User in Mobile Ad Hoc Networks

### 2.2 Limited Bandwidth

If all the connections follow entirely different routes the situation is similar to single connection scenario. But as shown in fig 1 (b) if more then one routes have to pass through a common neighboring node then one important consideration is whether or not the new neighboring node has sufficient bandwidth to support all the connections. If sufficient bandwidth is available then fine but what if it has bandwidth to support only some of the connections.

### 2.3 Proposed Solution for Multiple Connections

If it is not possible to re-establish all the routes simultaneously, then routes must be re-establish serially based on priority. If multiple connections re-constructed serially without considering the traffic type they are carrying then, real time traffic may experience more delay then the other traffic types. The routes carrying real time traffic must be given higher priority. The connections carrying delay tolerable traffic may be given lower priority. For example if a user is downloading a file using one connection and talking to someone using another connection, the second connection should be re-constructed before first one as it is carrying real time traffic.

When mobile node moves to new location and its neighboring nodes are changed, the routes for all the connections have to pass through the new neighbors. If routes for more then one connections have to pass through the same neighboring node as shown in the Fig 1 (b), an important consideration is whether or not the new neighboring node has sufficient bandwidth to support all the connections. In case new neighboring node do not have sufficient bandwidth to support all the connections, the source node may consider re-negotiation of the bandwidth or possible blocking of one or more connections especially the one with low priority and demanding a lot of bandwidth. Some connections should be dropped in order to allow other connection to continue. The connections should be dropped according to their priority and bandwidth requirement. It is better to drop a single connection requiring low priority with large amount of bandwidth instead of dropping several connections which require low bandwidth.

## 3 PROPOSED ALGORITHM: NEW

As demonstrated in this document, the numbering for sections upper case Arabic numerals, then upper case Arabic numerals, separated by periods. Initial paragraphs after the section title are not indented. Only the initial, introductory paragraph has a drop cap.

Primary objectives of any algorithm are, to discover the route between a pair of nodes, to manage the routing table information, and to maintain the route. Route discovery and management of the routing tables is similar whether it is single connection, or multiple connection environments. But route maintenance is quite different in both the cases. Proposed algorithm is concerned only with the route maintenance feature of the routing. A route maintenance procedure is proposed here for multiple connections environment with some necessary modification in AODV. The original AODV has been compared with the modified one proposed algorithm.

**1**. When a node moves, it initiates route discovery procedure for all the live connections concurrently or serially. In case connections are taken serially, high priority connections are handled first.

**2**. More than one connection may pass through the same intermediate node. If sufficient bandwidth is not available to support all the connections, required bandwidth is renegotiated. One or more connections are dropped to allow other connections to continue. Low priority connections with high bandwidth requirement are more eligible to be dropped.

**3**. If source node moves, a fresh path is discovered for every connection.

**4**. If an intermediate node moves out of the path, it sends route request (RREQ) message to the immediate downstream node (next node). The receiver of RREQ message starts route discovery procedure as source node.

**5**. When receiver of RREQ message fails to discover the path, it sends route failure message to the source node. Now source node starts route discovery process to find out entirely new route.

## 4 SIMULATION ENVIRONMENT

Simulation environment used for this study is ns-2 [10]. NS provides support for multi-hop wireless networks or MAC sub layer and for wireless environments. An on-demand routing protocol is used when the source has packets to send and the receiver can select the best route through various route selection criteria, such as those stated in AODV. For simplicity, the shortest path among multiple different paths is chosen to be the best route at the receiver in our simulation. Transmission ranges of nodes are assumed to be static, and both symmetric and asymmetric links are present in the network. Initially 50 nodes are scattered onto a 1000 x 1000 unit grid as shown in fig 2. Each node moves in random direction at a random velocity. To measure the packet delivery, both TCP (Transmission control packet) and UDP (User Datagram Packet) packets are used into the network from the source to the destination. Two most popular protocols AODV and DSR have been used for comparison.

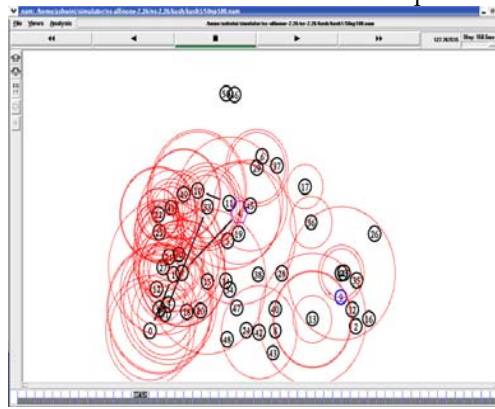

Fig 2: Simulation Scenario

Performance parameters used in this study are Packet Delivery Ratio, Average Message Overhead and throughput. Here Packet delivery can be defined as the fraction of successfully received packets, which survive while finding their destination. Average message overhead is defined as the ratio of total no. of message sent upon total no. of connection request and the rate of successfully transmitted data per second in the network during the simulation is called throughput.

## 5 RESULTS AND OBSERVATIONS

Results for the 'New' approach with 50 nodes using TCP connections have been shown in fig. 3(a). The area used is 1 Km × 1 Km and sources connected are 20-30. Another simulation was carried out using 100 nodes in the same scenario with 30-40 sources connected and fig. 3(b) show the results.

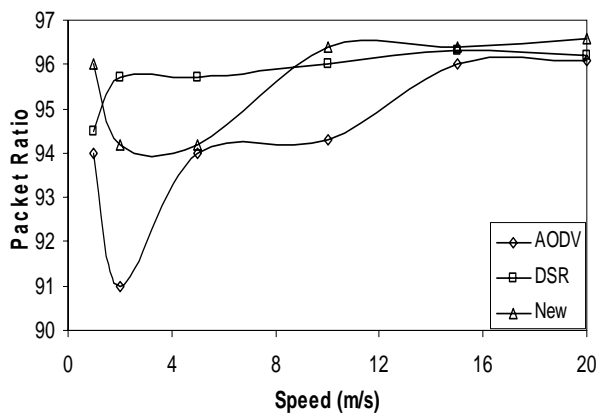

Fig. 3(a): packet delivery ratio with 50 nodes

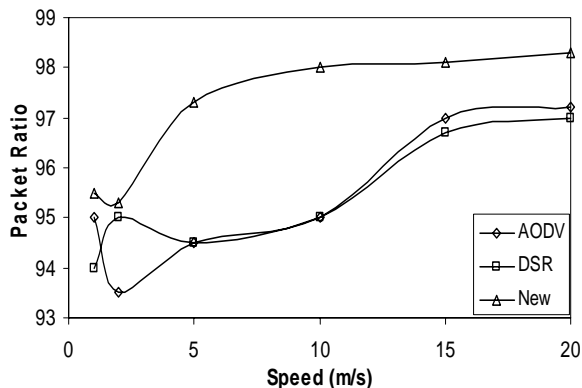

Fig. 3(b): packet delivery ratio with 100 nodes

NEW performs better at all speed except 2 m/s in a 50 nodes network as shown in figure 3(a) as compare to AODV and DSR both. Performance suffers a bit at higher speed in denser medium of 100 nodes as shown in figure 3 (b).. The reason is that keeping cache for such a large network demand more storage and in turn slows packet delivery rate. DSR is able to deliver more than 94% packets all the time. AODV improves in denser mediums as it is able to support more packets. It overpowers DSR at high speed of 15 to 20 meters per second and trend is true even at higher speeds. NEW has been the best in dense mediums, showing almost same performance at all speeds. In case of NEW delivery ratio was nearing 98% even at higher speeds 10 to 20 m/sec. NEW performs better in denser medium (100 nodes), as more nodes are available for route selection, so better bandwidth distribution for different data traffic is possible.

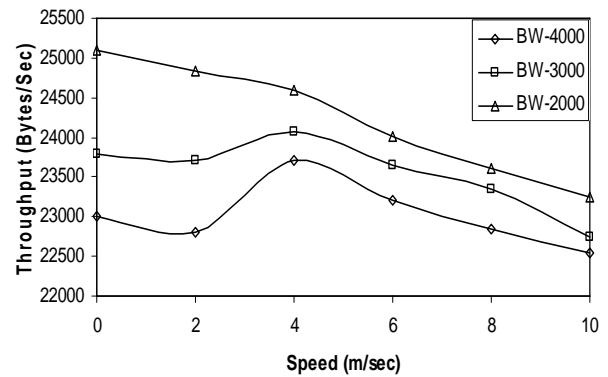

Fig. 4: Throughput Vs Mobility for New

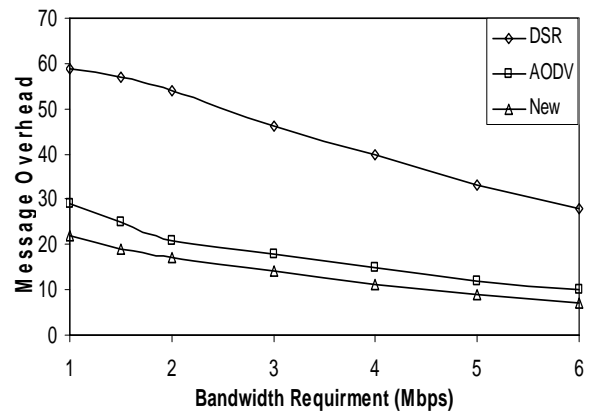

Fig. 5: Message overhead per connection request

In Figure 4, an analysis of throughput against mobility is shown for NEW with 100 nodes and a bandwidth demanded for different connections, while the value of the demanded QoS parameter (bandwidth) changes from 2000 to 4000. It is evident from the figure that with the high speed more than 4 m/sec the value for throughput for all demanded QoS path decreases for NEW. Because frequent route failures due to mobility causes more route reconstructions which leads to more packet loss. It can also observe from the figure that high bandwidth demanded QoS connections has high throughput all the time at different speeds.

At last, average message overheads per connection are analyzed for a 100 nodes network with a bandwidth requirement of 1-6 Mbps with DSR, AODV and NEW. The results for all the protocols are shown in Figure 5. It is observed from the figure that NEW and AODV bears significantly less overheads then that of DSR. It is also observed that NEW with multiple connections has slight fewer overheads even as compare to AODV. It is evident from the fig. 6 that even at a higher input bandwidth requirement, in AODV message overheads per connection request is less than 10 while for the same situation DSR incurred around 35 message overheads per connection request. This significant reduction in message overheads is achieved due to limited broadcast nature of AODV.

## 6 CONCLUSION

In this paper the kinds of trouble that can arise in maintaining multiple connections in adhoc networks have been described clearly which are not pertinent in case of single connections. This is a well motivated problem and current literature does not provide enough material to learn about multiple connections in adhoc networks. This paper provides a novel solution to maintain multiple connections in an adhoc network. The algorithm mentioned in the paper has been carefully studied by the use of simulation tests. The tests provide evidence to show that the proposed algorithm 'NEW' performs better than existing methodologies. This paper assures that if such issues will be taken in to account for adhoc networks, these would truly be helpful to other researchers in the area. This scheme is of its own kind, which not only takes care of multiple connections in adhoc routing but also maintains them with efficient bandwidth utilization while considering different traffic types. The scheme can be incorporated into any ad hoc on-demand unicast routing protocol to improve reliable packet delivery in the face of node movements and route breaks. Alternate routes are utilized only when data packets cannot be delivered through the primary route. As a case study, the proposed scheme has been applied to AODV and it was observed that the performance improved. Simulation results indicated that the technique provides robustness to mobility and enhances protocol performance. Study is going on currently investigating ways to make this new protocol scheme robust to traffic load.

The process of checking the protocol scheme is on for more sparse mediums and real life scenarios and also for other metrics like Path optimality, Link layer overhead.

The proposal is also on to check this protocol for multicast routing.